\documentclass[a4paper,fleqn,usenatbib]{mnras}
\usepackage[T1]{fontenc}
\usepackage{ae,aecompl}
\usepackage{graphicx}	
\usepackage{amsmath}	
\usepackage{amssymb}	
\usepackage{wasysym}
\usepackage{lscape}
\usepackage{natbib}
\usepackage[varg]{txfonts}
\usepackage{url}
\usepackage{xspace}
\usepackage{xcolor}
\usepackage{longtable}
\usepackage{pdflscape}	
\bibpunct{(}{)}{;}{a}{}{,}
\pdfoutput=1

\newcounter{Rco}
\newcommand{\Ionst}[1]{\setcounter{Rco}{#1}\Roman{Rco}}
\newcommand{\Ion}[2]{\mbox{#1\,{\scriptsize\Ionst{#2}}}}
\newcommand{\Ionw}[3]{\mbox{#1\,{\scriptsize\Ionst{#2}}~$\lambda\,#3$\,\AA}}

\newcommand{\Ionww}[3]{\mbox{#1\,{\scriptsize\Ionst{#2}}~$\lambda\lambda\,#3$\,\AA}}

\newcommand{\logg}{\mbox{$\log g$}\xspace}
\newcommand{\loggw}[1]{\mbox{$\log g\hspace{-0.5mm} =\hspace{-0.5mm}  #1$}}

\newcommand{\Teff}{\mbox{$T_\mathrm{eff}$}\xspace}
\newcommand{\Teffw}[1]{\mbox{$\Teff\hspace{-0.5mm} =\hspace{-0.5mm} #1 \,\mathrm{K}$}}

\newcommand{\Lsol}{$L_\odot$}
\newcommand{\Msol}{$M_\odot$}
\newcommand{\Rsol}{$R_\odot$}

\newcommand{\wda}{\mbox{J0536+5448}\xspace}
\newcommand{\wdb}{\mbox{J2311+2929}\xspace}

\title[Discovery of two bright DO-type white dwarfs]{Discovery of two bright DO-type white dwarfs
\thanks{Based on observations collected at the German-Spanish Astronomical Center, Calar Alto, jointly operated by 
           	the Max-Planck-Institut f\"{u}r Astronomie Heidelberg and the Instituto de Astrof\'{i}sica de Andaluc\'{i}a (CSIC).}
\thanks{Based on observations made with the William Herschel Telescope and the Isaac Newton Telescope operated on the island of La Palma by the Isaac Newton Group of Telescopes in the Spanish Observatorio del Roque de los Muchachos of the Instituto de Astrof\'{i}sica de Canarias.}
}

\author[Reindl et al.]{
Nicole Reindl,$^{1}$\thanks{E-mail: nr152@le.ac.uk}
S. Geier,$^{2}$
R.~H.~\O stensen$^{3}$
\\
$^{1}$Department of Physics and Astronomy, University of Leicester, University Road, Leicester LE1 7RH, UK\\
$^{2}$Institute for Physics and Astronomy, University of Potsdam, Karl-Liebknecht-Str. 24/25, D-14476 Potsdam, Germany\\
$^{3}$Department of Physics, Astronomy and Materials Science, Missouri State University, Springfield, MO 65897, USA\\
}
\date{Accepted XXX. Received YYY; in original form ZZZ}

\pubyear{2018}

\begin{document}
\label{firstpage}
\pagerange{\pageref{firstpage}--\pageref{lastpage}}
\maketitle

\begin{abstract}
We discovered two bright DO-type white dwarfs, \mbox{GALEX\,J053628.3+544854} (\wda) and \mbox{GALEX\,231128.0+292935}(\wdb), which rank among the eight brightest DO-type white dwarfs known. Our non-LTE model atmosphere analysis reveals effective temperatures and surface gravities of \Teffw{80000\pm4600} and \loggw{8.25\pm0.15} for \wda and \Teffw{69400\pm900} and \loggw{7.80\pm0.06} for \wdb. The latter shows a significant amount of carbon in its atmosphere ($C=0.003^{+0.005}_{-0.002}$, by mass), while for \wda we could derive only an upper limit of $C<0.003$. Furthermore, we calculated spectroscopic distances for the two stars and found a good agreement with the distances derived from the Gaia parallaxes.
\end{abstract}



\section{Introduction}
\label{sect:intro}

About every fifth intermediate mass star will turn into a star with a H-deficient atmosphere in its late stage of stellar evolution. 
There are various scenarios that can produce H-deficient stars, such as (very) late thermal pulses ( e.g., \citealt{Iben1983, Althaus2005}), 
the merger of two white dwarfs \citep{Zhangetal2012a, Zhangetal2012b}, or the late hot flasher scenario \citep{Millerbertolamietal2008}.
Once these H-deficient stars enter the white dwarf cooling sequence, their atmospheres are dominated by He. The earliest type of these 
He-dominated white dwarfs corresponds to the DO spectral type. These stars have effective temperatures (\Teff) between $45\,000$ and $200\,000$\,K, 
with the hot DO white dwarfs showing strong \Ion{He}{2} lines, whereas in the spectra of cool DO white dwarfs \Ion{He}{1} lines can also be seen 
\citep{Sion2011}.\\
DO-type white dwarfs cover a huge luminosity interval from up to $10^4$ times the solar luminosity, when they have just entered the white 
dwarf cooling sequence, down to only 0.1\,\Lsol when reaching to the DB-spectroscopic region. This evolutionary phase is quite short-lived 
($\approx 2$\, million years, \citealt{Althausetal2009}), thus less than 100 DO white dwarfs are currently known.\\
The detection of more of these stars is important for the construction of a reliable hot end of the white dwarf luminosity function (WDLF).
Its hot end suffers form low number statistics, i.e., the high luminosity bins ($- 0.5 < M_{\mathrm{bol}} < 4.0$, corresponding to 
white dwarfs with \Teff $> 50000$\,K) of the hot WDLF of \cite{Krzesinski2009} contain only one to 29 objects each. Since the shape of the 
WDLF quantitatively reflects the cooling behaviour of the white dwarf population as a whole, it is an excellent tool for constraining 
the properties of particles emitted by the white dwarfs. The cooling process of hot and luminous white dwarfs is dominated by the radiation 
of neutrinos, thus, the shape of the hot end of the WDLF strongly constrains the magnetic dipole moment of the neutrino \cite{MillerBertolami2014b}. 
Moreover, it offers the opportunity to check for the possible existence of the axion, a proposed but not yet detected type of weakly interacting 
particles \citep{MillerBertolami2014a}.\\ 
DO-type white dwarfs with \Teffw{58000-85000} are found to display high abundances of trans-iron elements \citep{Hoyeretal2018a, Hoyeretal2018b}. Therefore, high resolution ultra-violet (UV) spectra of DO white dwarfs, in which a wealth of metal lines can be detected, enables us to 
study the chemical evolution of these stars and to put constraints on possible progenitors and progeny. Furthermore, hot white dwarfs 
serve as stellar laboratories to derive atomic data for highly ionized species of trans-iron elements \citep{Rauch2012, Rauch2014a, Rauch2014b, Rauchetal2015a, Rauchetal2015b, Rauchetal2016, Rauchetal2017a, Rauchetal2017b}.\\
Last but not least, hot white white dwarfs have recently found a new use in the fundamental physics community. They potentially 
allow us to directly observe variations in fundamental constants, i.e., the fine structure constant $\alpha$, at locations of high 
gravitational potential. Such a variation manifests itself as shifts in the observed wavelengths of absorption lines, when compared to laboratory 
wavelengths. This effect is larger for metals with more protons (higher atomic number) and higher ionization energy. Thus, \Ion{Fe}{5} and 
\Ion{Ni}{5} that can be found in high-resolution UV spectra of hot white dwarfs are particularly well suited for this purpose 
\citep{Berengutetal2013, Prevaletal2015, Bainbridgeetal2017}.\\
The Hubble Space Telescope (HST) is currently the only telescope that can be used to obtain UV spectra for the above mentioned proposes. 
Thus, bright white dwarfs ($V<16$\,mag) are a requisite to gain a high signal to noise (S/N), high-resolution UV spectrum, necessary to 
resolve the weak metal lines in hot white dwarfs.\\ 
In this letter we report the discovery of two such bright DO white dwarfs, \mbox{GALEX\,J053628.3+544854} and 
\mbox{GALEX\,231128.0+292935} (hereafter \wda and \wdb). Since most of the DO white dwarfs have been discovered within the 
Solan Digital Sky Survey, the majority of DO type white dwarfs are significantly fainter (mean V magnitude is $17.3$\,mag). The 
two stars presented here have V magnitudes of $15.186\pm0.062$ (\wda) and $15.491\pm0.040$ (\wdb, \citealt{Hendenetal2016}) and therefore 
they rank among the eight brightest DO-type white dwarfs known. The only six DO white dwarf brighter than those two are HZ\,21, 
HS\,0111+0012, RE\,0503$-$289, PG\,0109+111, PG\,0038+199, and PG\,1034+001. 

\begin{figure*}
 \centering
  \includegraphics[width=\textwidth, scale=0.80]{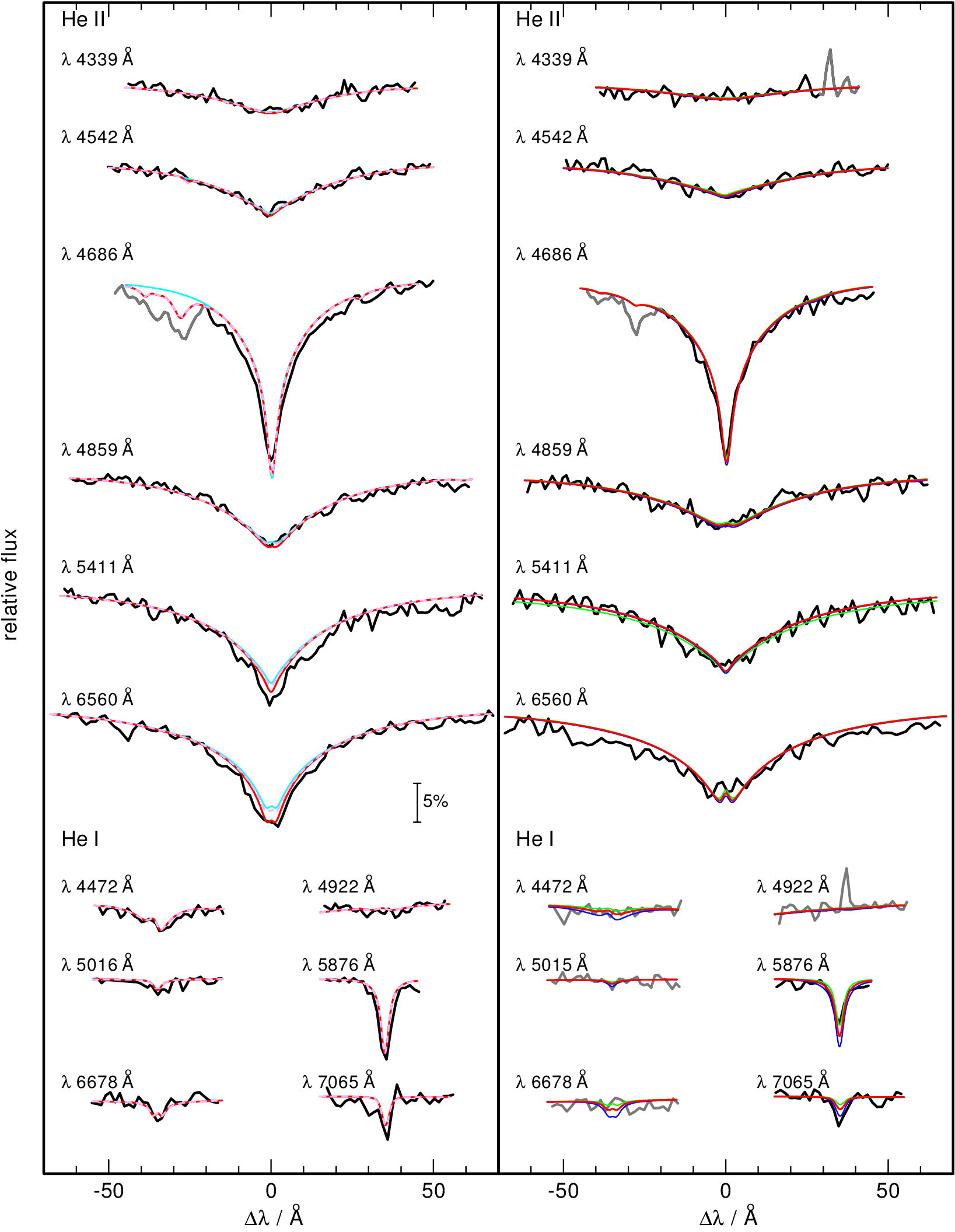}
  \vspace*{-5mm}
  \caption{Coadded and normalized TWIN spectra (black) of \wdb (left) and \wda (right). In the left panel the impact of C, N, and O on the theoretical lines profiles is demonstrated (light blue: pure He model, dashed pink: HeC model, red: HeCNO model). In the right panel models with different \Teff\ are shown (red: 80000\,kK, blue: 75400\,kK, green 84600\,kK). Regions excluded from the $\chi^2$ fit  are shown in grey. The vertical bar indicates 5\,\% of the continuum flux.}
  \label{fig:twin}
\end{figure*} 

\section{Spectroscopic observations}
\label{sect:obs}

We observed \wda and \wdb during surveys exploring candidate sdB stars for the Kepler mission, selected from the GALEX survey.
In August 2010 \wda was observed twice at the William Herschel Telescope using the Intermediate dispersion Spectrograph 
and Imaging System (ISIS) with the R600B grating ($R\approx 4000$), covering the wavelength range $3500-5200\,\AA$. \wda 
was again observed at the Isaac Newton Telescope (INT) February 2011 using the Intermediate Dispersion 
Spectrograph (IDS) with the R400B grating ($R\approx 1500$) covering the wavelength range of $3500-9000\,\AA$. 
\wdb was observed with the same set-up at the INT in July 2013. Based on these spectra we assigned the spectral type DO.\\
To obtain high S/N spectra suitable for a precise spectral analysis, we performed spectroscopic follow-up in October/November 2014 
at the Calar Alto $3.5$\,m telescope (ProgID H14-3.5-022) using the TWIN spectrograph and a slit width of 1.2 acrsec. 
We used the grating No. T08 for the blue channel and No. T04 for the red channel (dispersion 72 $\AA/$nm, covering the 
wavelength range $3500-7300\,\AA$). The resolution of the spectra is $1.8\,\AA$.
We obtained four spectra of \wda and seven spectra of \wdb with exposure times of 20\,minutes to 
achieve a S/N $> 100$. The spectra were taken in three consecutive nights. After each spectrum, we required ThAr wavelength calibration. 
The data reduction was done by using IRAF\footnote{IRAF is distributed by the National Optical Astronomy Observatory, which is operated 
by the Associated Universities for Research in Astronomy, Inc., under cooperative agreement with the National Science Foundation.}. 
We did not perform flux calibration to our data.\\
The co-added TWIN spectrum of \wda reaches a S/N (interpreted as mean/standard deviation, measured in the 
region $6000-6250\,\AA$) of 120 and displays \Ionww{He}{2}{4339, 4542, 4686, 4859, 5411, 6560}, \Ionww{He}{1}{5876, 7065}. 
The co-added TWIN spectrum of \wdb reaches a S/N of 200 and additionally shows lines of \Ionww{He}{1}{4472, 4922, 5016, 6678} and 
\Ionww{C}{4}{4646-4660, 5814, 5803}. To check for radial velocity variations, we measured the radial velocities of the 
identified lines by fitting a set of mathematical functions (Gaussians, Lorentzians, and polynomials) to the identified spectral 
lines using SPAS (Spectrum Plotting and Analysing Suite, \citealt{Hirsch2009}). None of the two stars displays any radial 
velocity variations larger than $20$\,km/s, corresponding to the standard deviation of the individually measured radial 
velocities.

\section{Spectral analysis}
\label{sect:ana}

\subsection{Effective temperatures and surface gravities}
\label{subsect:eff}

We derived the effective temperatures and surface gravities of the two stars by means of a $\chi^2$ minimization technique with 
SPAS, which is based on the FITSB2 routine \citep{Napiwotzki1999}. The fit is based on the co-added TWIN spectra of the two stars 
and the model grid of \cite{Reindletal2014c}, which was calculated with the T{\"u}bingen non-LTE Model-Atmosphere Package 
(TMAP, \citealt{werneretal2003, tmap2012, rauchdeetjen2003}). To account for the spectral resolution of the TWIN observations, 
the synthetic spectra were convolved with Gaussians ($FWHM = 1.8$\AA). We fitted the \Ionww{He}{2}{4339, 4542, 4686, 4859, 5411, 6560}, 
\Ionww{He}{1}{5876, 7065}, and for \wdb additionally \Ionww{He}{1}{4472, 4922, 5016, 6527}. For \wda we derived \Teffw{80000 \pm 4600} 
and \loggw{8.25 \pm 0.15}, and for \wdb we found \Teffw{69400 \pm 900} and \loggw{7.77 \pm 0.06}. The errors represent the 
statistical one sigma uncertainty on \Teff and the three sigma uncertainty on \logg.\\
In Fig.~\ref{fig:twin} we show the fits to the He lines of the co-added TWIN spectra of the two stars. For \wda (left panel of Fig.~\ref{fig:twin}), 
the \Ion{He}{2} lines are nicely reproduced by the best fit model (red), but the theoretical \Ionw{He}{1}{7065} is too shallow and 
the \Ionw{He}{1}{5876} too deep. We also show in this figure the models which are 4600\,K (error from the $\chi^2$ fit) hotter (green) 
and 4600\,K cooler (blue). While the cooler model reproduces best the \Ionw{He}{1}{7065}, the hotter one matches 
\Ionw{He}{1}{5876}. For \wdb (left panel of Fig.~\ref{fig:twin}), the \Ion{He}{1} lines are reproduced well, but the line cores of 
\Ionww{He}{2}{5411, 6560} are too shallow in the pure He model (light blue).\\
As a next step, we calculated TMAP model grids, which also include opacities of C, N, and O. The statistics of our model atoms used for 
the model atmosphere calculations are summarized in Table~\ref{tab:modelatom}.    
The addition of C in the models does not affect the theoretical \Ion{He}{1} and \Ion{He}{2} (see also \citealt{Reindletal2014c}). We also 
demonstrate this in left panel of Fig.~\ref{fig:twin}, where a HeC model (dashed pink) is compared to the pure He model (light blue), both 
with \Teffw{69400} and \loggw{7.77}. Comparing these models to our HeCNO models, we find that the line cores of 
\Ionww{He}{2}{5411, 6560} become deeper for increasing abundances of N and O. In case of \wdb this indeed solves the problem of the observed 
too deep \Ionww{He}{2}{5411, 6560} lines (red line in the left panel of Fig.~\ref{fig:twin} represents the HeCNO model). We then repeated 
the $\chi^2$ fit for \wdb based on the HeCNO models. We found that \Teff\ is not affected by the inclusion of CNO (the derived \Teff\ is 
mainly a result of the strengths of the \Ion{He}{1} lines, which are not affected by the inclusion of CNO) and that the derived surface gravity 
increases only by 0.03\,dex. We will therefore adopt in following, one sigma errors on \Teff\ and three sigma errors on \logg. 

\subsection{Carbon}

\begin{figure}
\includegraphics[width=\columnwidth]{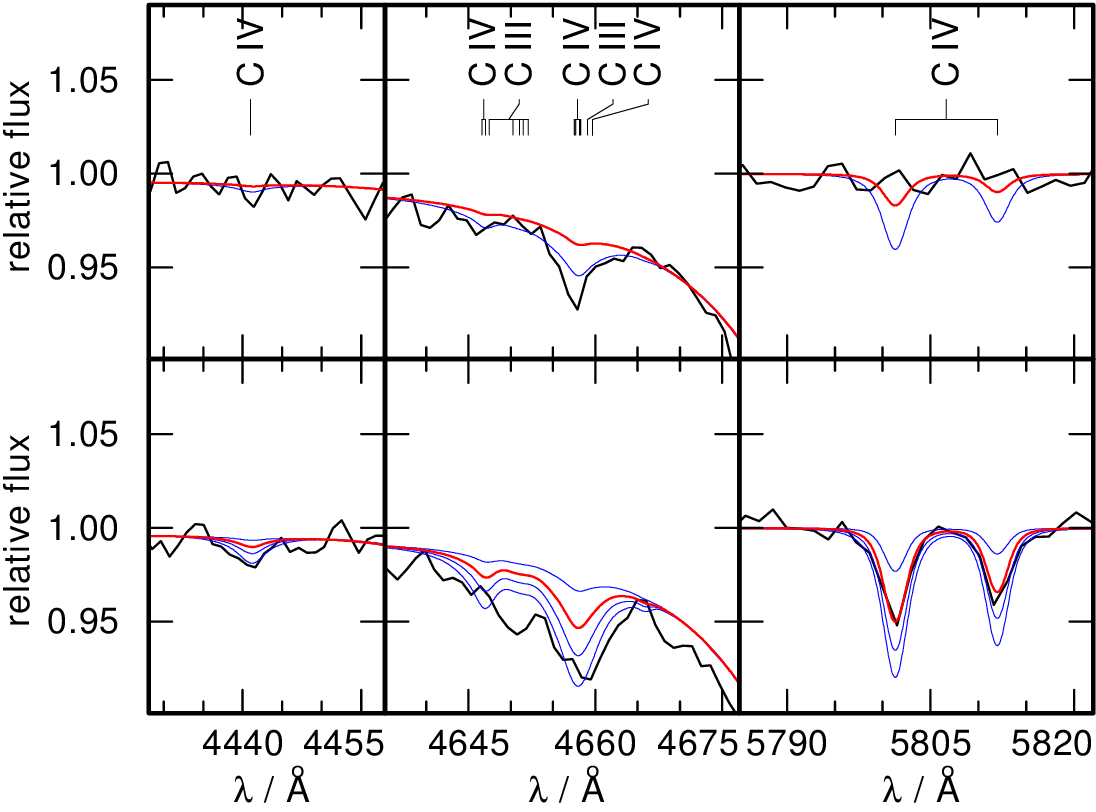}
\vspace*{-5mm}
\caption{Determination of the C abundance of \wdb (bottom panel) and the upper limit on the C abundance for \wda (upper panel). The red line indicates the (upper limit of) C abundance as derived from \Ionww{C}{4}{5801, 5812} in the TWIN spectra. C abundances in the models for \wdb are 0.001, 0.003, 0.005, and 0.008 and for \wda 0.001 and 0.003 (mass fractions).}
\label{fig:civ}
\end{figure}

For \wdb the observed \Ionww{C}{4}{5802, 5012} lines are well reproduced for $C=0.003$ (mass fraction), but not the absorption at 
$4646-4660\,\AA$, which suggest $C=0.008$ (Fig.~\ref{fig:civ}. Our models predict also \Ionww{C}{4}{4441}. There is small 
absorption at this wavelength in the spectrum of \wdb, which however lies within the noise level of the spectrum, leaving the 
identification of this line uncertain. \wda does not show \Ionww{C}{4}{5802, 5012} suggesting $C<0.001$. The TWIN spectrum of this 
star, however, does show an absorption in the $4646-4660\,\AA$ region, which could be due to \Ionww{C}{4}{4646-4660} and suggests 
$C=0.003$. The IDS spectra of both stars are to noisy to allow the identification of \Ionww{C}{4}{5802, 5012}, but both show the 
absorption at $4646-4660\,\AA$, too.\\
This problem cannot be overcome with different combinations of  \Teff\ and \logg. An increase of \Teff\ or a decrease \logg\
affects, however, \Ionww{C}{4}{4646-4660} and \Ionww{C}{4}{5802, 5012} in the same manner. For example, if \Teff\ is increased by 
5000\,K, or \logg\ decreased by 0.5\,dex, this has a similar effect on those C lines as an increase of the abundance by 0.001. 
A decrease of \Teff\ by 5000\,K, decreases the line strengths of all the \Ion{C}{4} lines only minimally 
(the line strength of \Ionww{C}{3}{4650-4652} increases slightly in the cooler model for \wdb). Thus, a lower \Teff\ can also not 
account for the discrepancy of C abundance derived from the different \Ion{C}{4} lines.
We emphasize that both the \Ionww{C}{4}{4646-4660} complex as well as the \Ionww{C}{4}{5802, 5012} lines form at a similar 
part of the atmosphere ($\log m \approx -2$, where $m$ is column mass that is measured from the outer boundary of the model atmospheres.), 
which means that it is unlikely that this problem could be solved by the inclusion of further metal opacities that would impact 
the temperature structure of the model atmosphere. So far, this problem has not been reported in the analysis of any other DO-type 
white dwarf. However, for some PG\,1159 also only a poor simultaneous fit of the \Ionww{C}{4}{4646-4660} and the \Ionww{C}{4}{5802, 5012}
lines is achieved. Similar to \wda, the spectrum of the PG\,1159 star HS\,1517+7403 does show strong \Ionww{C}{4}{4646-4660} lines while 
\Ionww{C}{4}{5802, 5012} cannot be detected (Fig.\,5 of \citealt{Dreizler1998}). Another example is PG\,0122+200 for which \cite{WernerRauch2014} 
could not precisely reproduce the observed central emission of \Ionww{C}{4}{4659-4660}. Thus, an inaccuracy in the atomic 
data could be origin of this problem.\\
Another possible solution is a blend with a ultra-high excitation (uhe) absorption line, i.e., \Ionw{O}{8}{4658}. This line is seen in 
about every tenth DO-type white dwarf \citep{Werneretal1995, Dreizleretal1995, Werneretal2014, Reindletal2014c}, which are commonly 
refereed to as hot-wind white dwarfs. The reason for the occurrence of these lines is not understood, but a photospheric origin can 
be excluded. On the other hand, the spectra of the hot wind white dwarfs are always accompanied by too broad and deep 
\Ion{He}{2} lines, which is not the case for our two stars. Also, should the (additional) line absorption around \Ionww{C}{4}{4646-4660} 
stem from an uhe line it would be the first time that only this uhe line is observed.

\subsection{Nitrogen and Oxygen}

In our HeCNO models \Ionw{N}{4}{4058} is the strongest of all nitrogen lines, followed by \Ionw{N}{4}{6381}. In the hotter models (\Teffw{84600}) for 
\wda, also the \Ionww{N}{5}{4604, 4620} doublet becomes slightly apparent. However, none of these lines are observed in the spectra. We 
derive an upper limit of $N<0.005$ for both stars. The strongest oxygen lines in our models are \Ionw{O}{3}{3962} followed by \Ionw{O}{4}{3737}. 
Those lines are not visible in the spectra of our stars. We determine an upper limit of $O<0.01$ for both stars. We summarize the atmospheric 
parameters of the two stars in Table~\ref{tab:parameters}.

\begin{table}
\caption{Statistics of our model atoms.}
\begin{tabular}{lrrr}
\hline\hline
\noalign{\smallskip}
       & Ion         & NLTE levels   & Lines \\  
\hline
He & {\sc i}   &     29 &     69            \\
   & {\sc ii}  &     20 &    190            \\
   & {\sc iii} &      1 &     $-$           \\
\noalign{\smallskip}
C  & {\sc ii}  &     16 &     37            \\
   & {\sc iii} &     67 &    101            \\
   & {\sc iv}  &     54 &    295            \\
   & {\sc v}   &      1 &     0             \\
\noalign{\smallskip}
N  & {\sc iii} &     13 &    24            \\
   & {\sc iv}  &     16 &    30            \\
   & {\sc v}   &     14 &    35            \\
   & {\sc vi}  &      1 &     0            \\
\noalign{\smallskip}
O  & {\sc iii} &     11 &     9            \\
   & {\sc iv}  &     18 &    39            \\
   & {\sc v}   &     17 &    25           \\
   & {\sc vi}  &     14 &    33            \\
   & {\sc vii} &      1 &      0           \\
\noalign{\smallskip}
\hline
\end{tabular}
\label{tab:modelatom}
\end{table}

\begin{table}
\caption{Atmospheric parameters of the two stars. Abundances are given in mass fractions.}
\begin{tabular}{lll}
\hline\hline
\noalign{\smallskip}
                & \wda   & \wdb \\  
\hline
\noalign{\smallskip}             
\Teff\ [K]             & $80000 \pm 4600$ & $69400 \pm 900$ \\   
\logg\ [cm\,s$^{-2}$]  & $8.25 \pm 0.15$  & $7.80  \pm 0.06$ \\
$C$                    & $< 0.003 $       & $0.003^{+0.005}_{-0.002}$ \\       
$N$                    & $< 0.005 $       & $< 0.005 $ \\ 
$O$                    & $< 0.01 $       & $< 0.01 $ \\ 
\noalign{\smallskip}
\hline
\end{tabular}
\label{tab:parameters}
\end{table}

\section{Masses, radii, and distances} 
\label{sect:masses} 

\begin{figure}
\includegraphics[width=\columnwidth]{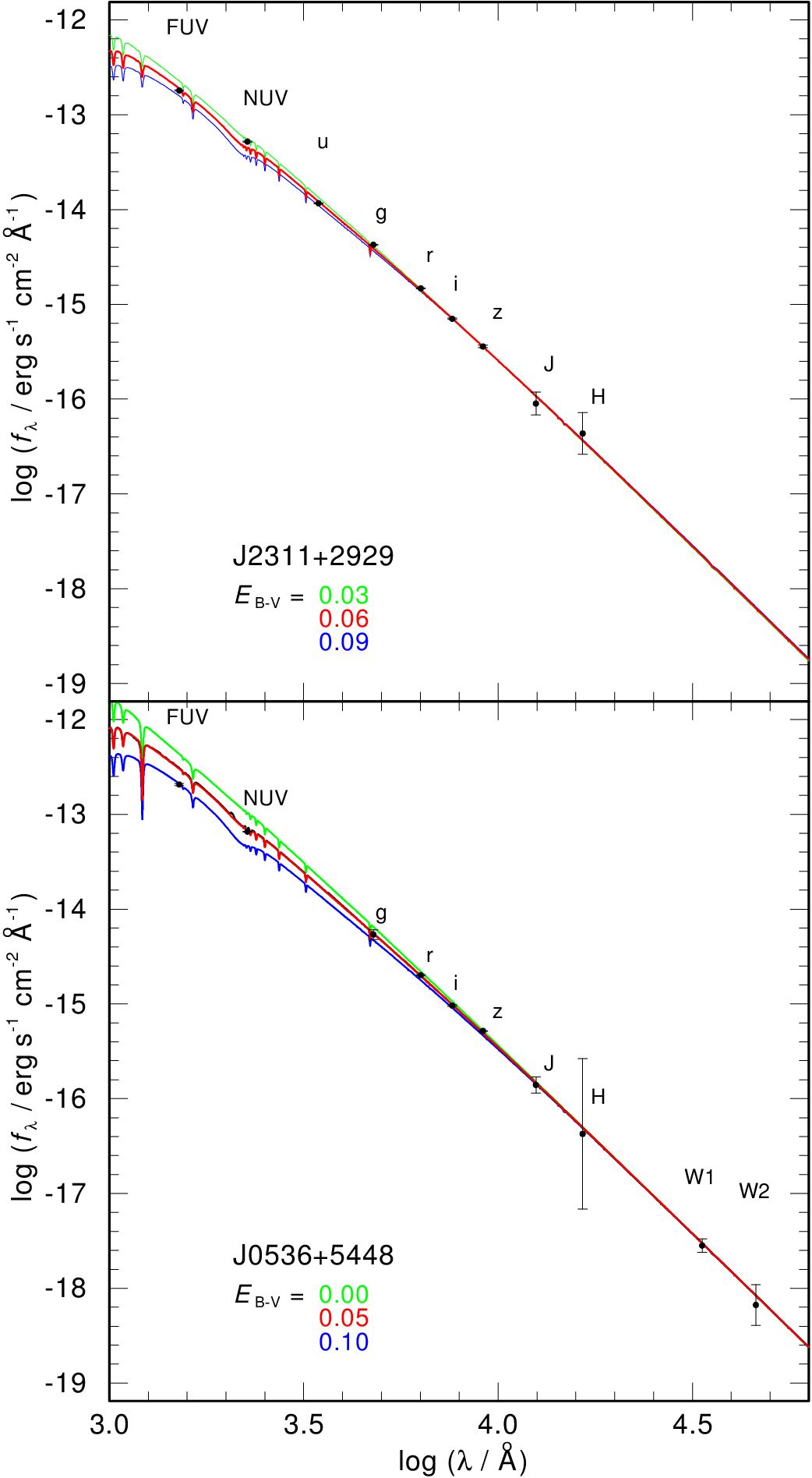}
\vspace*{-5mm}
\caption{Determination of the reddening of \wdb (upper panel) and \wda (lower panel).}
\label{fig:ebv}
\end{figure}

\wda and  \wdb are included in Gaia data release 2 \citep{Gaia2016, Gaia2018} and they have parallaxes of 
$\pi = 4.488696620299169\pm 0.08003018105955109$ and $3.84995585215932\pm 0.06101563607521435$, respectively. 
An additional systematic error of the parallax due to uncertainties on the parallax zero-point and spatial correlations, 
of 0.1\,mas is recommended by the Gaia team (see also \citealt{Zinnetal2018}). The resulting distances and height above 
or below the Galactic plane, $z$, are listed in Table~\ref{tab:parametersb}.\\
By calculating the spectroscopic distance, we can now check for the consistency with the derived atmospheric parameters of the two 
stars. For that we firstly calculated absolute synthetic $V$ band magnitude, $M_V$, according to \cite{HolbergBergeron2006} 
using $$M_V = - 2.5 \log \frac{\int R^2 H_{\lambda} S_V(\lambda)d\lambda}{\int S_V(\lambda)d\lambda} + 73.6484,$$
where $R$ is the radius of the star, $H_{\lambda}$ the Eddington flux of our model spectrum, and $S_V(\lambda)$ is the 
relative filter response modified for atmospheric transmission \citep{Cohenetal2003}. The spectroscopic distance can then be 
calculated via $m_{V_0}-M_V = 5\,\log (d[10pc])$, where $m_{V_0}$ is the observed $V$ band magnitude after corrected for 
extinction. The reddening $E_{B-V}$ of the stars was determined as follows. We obtained GALEX $FUV$ and $NUV$ magnitudes from 
\cite{Bianchi2014}, 2MASS $J$ and $H$ magnitudes from \cite{Lawrence2013}, $u$, $g$, $r$, $i$, and $z$ magnitudes for \wdb from 
\cite{Ahnetal2012}, $g$, $r$, $i$, $z$, $W1$, and $W2$ magnitudes for \wda from \cite{Chambersetal2016} and \cite{Cutri2014}, 
and converted them into fluxes. Next, we normalized our best fit model atmosphere flux to the $W1$ band flux for \wda and to the 
$z$ band flux for \wdb. Using the reddening law of \cite{fitzpatrick1999} we then applied different values of $E_{B-V}$ to the 
model spectrum until a good agreement with the multi-band photometry was achived (Fig.~\ref{fig:ebv}). For \wdb 
we found $E_{B-V}=0.06\pm0.03$ and $E_{B-V}=0.05\pm0.05$. At the Gaia distances the Bayestar17 3D dust map 
\citep{Greenetal2014, Greenetal2018} gives $E_{B-V}=0.06\pm0.02$ for \wdb and $E_{B-V}=0.12^{+0.11}_{-0.09}$ for \wda.
These values agree within the error limits, but for \wda $E_{B-V}>0.10$ can be excluded, else the observed UV magnitudes 
cannot be reproduced.\\
To calculate the error on the $M_V$, the uncertainties of the stellar radius and the Eddington flux need to be considered.
The uncertainty of the latter depends on \Teff, \logg, as well as the metal content of the model. In principle, a higher \Teff, 
lower \logg, and higher metal content (flux is redistributed from the UV towards longer wavelengths) leads to a higher value 
of $M_V$. 
The metal content of our stars is unknown, but to estimate the effect of metals on the Eddington flux in the $V$ band we obtained 
the model spectrum of the most metal-rich DO-type white dwarf RE\,0503-289 from the German Astrophysical Virtual Observatory service 
TheoSSA\footnote{\url{http://dc.zah.uni-heidelberg.de/theossa/q/web/form}}. Comparing the filter-weighted, integrated $V$-band 
fluxes of a pure He model with the one including metals (C, N, O, Si, P, S, Ca, Fe, Ni, Ge, Kr, Sn, Xe, Zn, and Ba), we find that the 
V-band flux would be underestimated by 4\%. Masses and radii of DO-type white dwarfs were obtained from very late thermal pulse 
evolutionary tracks from \citet[][Fig.\,\ref{fig:evo}]{Althausetal2009}. The uncertainty of the radius of the star is determined 
by the errors on \Teff\ and \logg. The lack of the knowledge of the envelope mass adds an additional error of 
about 5\%.\\
The derived spectroscopic distances are listed in  Table~\ref{tab:parameters} and agree well with the Gaia distances. This 
confirms the reliability of our spectral analysis and the theoretical evolutionary calculations. In turn, we can now also calculate the 
radii of the two DO white dwarfs from the Gaia parallaxes, and using the mass-radius relationship of \citep{Althausetal2009} also 
determine their masses. The results are listed in Table~\ref{tab:parametersb}. Note that the larger errors for \wda on the radii and 
masses derived from the Gaia data result from the larger uncertainty on \Teff\ and $E_{B-V}$.

\begin{table}
\setlength{\tabcolsep}{0.3em} 
\caption{Distances, height above/below the Galactic plane, masses, and radii as measured from spectroscopy and from Gaia parallaxes.}
\begin{tabular}{lrrrr}
\hline\hline
\noalign{\smallskip}
                &\multicolumn{2}{c}{\wda} &  \multicolumn{2}{c}{\wdb} \\  
                & Spec. & Gaia & Spec. & Gaia\\  
\hline
\noalign{\smallskip} 
$d$ [pc]    & $156^{+59}_{-38}$         & $223\pm9$                  & $235^{+44}_{-25}$          & $260\pm11$  \\
\noalign{\smallskip} 
$z$ [pc]    & $32^{+13}_{-7}$           & $46\pm1$                   & $-112\pm11$                & $-124\pm5$   \\
\noalign{\smallskip}               
$M$ [\Msol] & $0.820^{+0.080}_{-0.090}$ & $0.630^{+0.165}_{-0.080}$  & $0.597\pm0.020$ & $0.575\pm0.005$ \\
\noalign{\smallskip} 
$R$ [\Rsol] & $0.011^{+0.003}_{-0.001}$ & $0.011^{+0.002}_{-0.001}$  & $0.016\pm0.001$  & $0.0174\pm0.0002$\\
\noalign{\smallskip}
\hline
\end{tabular}
\label{tab:parametersb}
\end{table}

\section{Summary and conclusions} 
\label{sect:sum}

\begin{figure}
\includegraphics[width=\columnwidth]{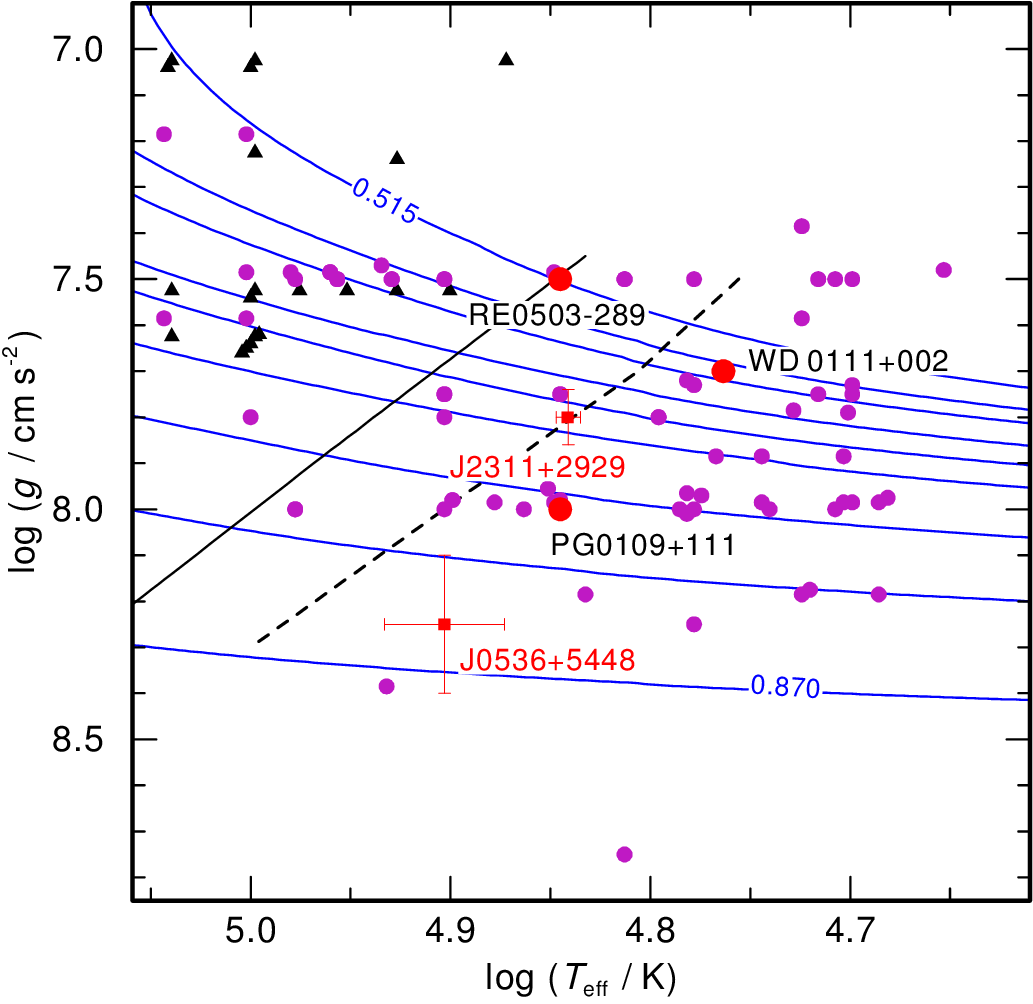}
\vspace*{-5mm}
\caption{Locations of the newly discovered DO white dwarfs (red squares) and other known DO white dwarfs (purple dots, \citealt{Reindletal2014c, Werneretal2014,  Huegelmeyeretal2006, dreizlerwerner1996}, red dots \citealt{Hoyeretal2018a, Hoyeretal2018b}) and PG\,1159 stars (black triangles, \citealt{Werneretal2014, Gianninas2010, wernerherwig2006, Schuh2008}) in the log \Teff\ -- \logg\ plane compared to very late thermal pulse evolutionary tracks (solid lines, the lowest and highest mass tracks are labeled with stellar masses, intermediate tracks correspond to 0.530, 0.542, 0.565, 0.584, 0.609, 0.644, and 0.741\,\Msol, respectively, \citealt{Althausetal2009}). The black lines indicate when the initial C abundance has dropped by a factor of two (solid line) and a factor of ten (dashed line, \citealt{UnglaubBues2000}).}
\label{fig:evo}
\end{figure}

We reported about the discovery of two bright DO-type white dwarfs, \wda and \wdb, which rank among the eight brightest
DO-type white dwarfs known. We performed a non-LTE spectral analysis of these two objects to derive their effective 
temperatures, surface gravities, C abundances (upper limit for \wda), and upper limits for N and O. 
For \wdb the observed line cores of \Ionww{He}{2}{5411, 6560} are too deep when compared to the best fit pure He model. 
We could solve this problem by including CNO in our model atmosphere calculations. In \wda this problem is not apparent, 
likely because this star has lower metal content. We found a discrepancy of C abundance derived from \Ionww{C}{4}{5802, 5012} 
and \Ionww{C}{4}{4646-4660}, as the latter two lines suggest a 0.002-0.003 lower C abundance. An inaccuracy of the atomic 
data or a blend with an uhe line or another hitherto unknown line could be possible reasons. No lines of N and O could be 
detected, which demonstrates that low resolution optical spectra are not suited to detect and derive abundances of those 
elements. The atmospheric parameter of the two stars are summarized in Table~\ref{tab:parameters}.\\
Furthermore, we have determined the reddening and spectroscopic distances of the two stars. We found that our derived 
values for $E_{B-V}$ agree within the error limits with those from the Bayestar17 3D dust map. We also find an a good 
agreement of the spectroscopic distances, masses, and radii with those derived from Gaia parallaxes 
(Table~\ref{tab:parametersb}). This confirms the reliability of our spectral analysis with the statistical one sigma 
uncertainty on \Teff\ and a three sigma error on \logg.\\ 
In Fig.~\ref{fig:evo} we show the locations of the newly discovered DO white dwarfs in the log \Teff\ -- \logg\ plane 
compared to very late thermal pulse evolutionary tracks of \cite{Althausetal2009}. The black lines indicate when the 
initial C abundance has dropped by a factor of two (solid line) and a factor of ten (dashed line) as predicted by
\cite{UnglaubBues2000}. This explain why \wdb still shows some C and \wda does not. Figure.~\ref{fig:evo} also shows the 
locations of the three DO white dwarfs, RE\,0503-298, WD\,0111+002 and PG\,0109+111, studied by 
\cite{Hoyeretal2018a, Hoyeretal2018b}. These stars are located close to the positions of our two bright DO white dwarfs, 
therefore it seems likely that \wda and \wdb display high abundances of trans-iron elements, too. Both stars are very bright 
in the UV (Fig.~\ref{fig:ebv}, thus, they are excellent objects for follow-up observations in the ultra-violet 
to search for trans-iron group elements.\\
Another interesting point that becomes obvious from Fig.~\ref{fig:evo}, is that \wda ranks amongst the most massive DO 
white dwarfs. Comparing the mass obtained form the Gaia parallax with the spectroscopic mass, we obtain a mass range of 
$0.730-0.795$\,\Msol. Because the theoretical variation of the fine structure constant $\alpha$ with gravity is expected to 
increase with the compactness ($M/R$) of an object, the detection of \Ion{Fe}{5} and \Ion{Ni}{5} lines in the spectrum of \wda would 
be especially interesting. We note that the compactness of \wda could exceed the ones in the sample of \cite{Bainbridgeetal2017} 
by a factor of $1.4$, suggesting that if $\alpha$ exhibits a gravitational dependence the effect could be notably larger than 
the previous measurements. For example, if we assume for demonstrational purposes that the result from \cite{Berengutetal2013} for 
\Ion{Fe}{5} is correct and can be generalised to \wda then the variation of the fine structure constant could be as large as 
$1 \times 10^{-4}$. This is $2.5$ times larger the \cite{Berengutetal2013} estimate for \mbox{G191$-$B2B} using \Ion{Fe}{5} 
($4.2 \pm 1.6 \times 10^{-5}$). A successful detection of such a variation would be the first direct measurement of a 
gravitational field effect on a bare constant of nature.

\section*{Acknowledgements}
NR is supported by a research fellowship of the Royal Commission for the Exhibition of 1851.
SG acknowledges funding by the Heisenberg program of the Deutsche Forschungsgemeinschaft under grant GE 2506/8-1.
This research has made use of NASA's Astrophysics Data System, the VizieR catalogue access tool and the SIMBAD 
data base operated at CDS, Strasbourg, France. The TheoSSA service (\url{http://dc.g-vo.org/theossa}) and TMAD 
service ((\url{http://astro.uni-tuebingen.de/~TMAD/}) used to retrieve theoretical spectra and model atoms for 
this paper were constructed as part of the activities of the German Astrophysical Virtual Observatory.
This work has made use of data from the European Space Agency (ESA) mission {\it Gaia} 
(\url{https://www.cosmos.esa.int/gaia}), processed by the {\it Gaia} Data Processing and Analysis Consortium 
(DPAC, \url{https://www.cosmos.esa.int/web/gaia/dpac/consortium}). Funding for the DPAC has been provided by 
national institutions, in particular the institutions participating in the {\it Gaia} Multilateral Agreement.



\bibliographystyle{mnras}
\bibliography{hotwind}


\bsp	
\label{lastpage}
\end{document}